# Ontology-based Context Aware Recommender System Application for Tourism.

Vitor T. Camacho[1], José Cruz[2]

[1] PhD, vitor.camacho@syone.com, R&D Data Science, Syone.

[2] MSc, jose.cruz@syone.com, R&D Data Science, Syone.

*Abstract*

In this work a novel recommender system (RS) for Tourism is presented. The RS is context aware as is now the rule in the state-of-the-art for recommender systems and works on top of a tourism ontology which is used to group the different items being offered. The presented RS mixes different types of recommenders creating an ensemble which changes on the basis of the RS's maturity. Starting from simple content-based recommendations and iteratively adding popularity, demographic and collaborative filtering methods as rating density and user cardinality increases. The result is a RS that mutates during its lifetime and uses a tourism ontology and natural language processing (NLP) to correctly bin the items to specific item categories and meta categories in the ontology. This item classification facilitates the association between user preferences and items, as well as allowing to better classify and group the items being offered, which in turn is particularly useful for context-aware filtering.

## 1 Introduction

This work presents a novel recommender system (RS) approach, which builds on context awareness, domain ontology and different types of recommenders that enter the process at different stages of maturity. From simple recommenders that are less prone to cold-start issues to more complex and powerful recommenders which struggle quite a bit with initial lack of data. At the final stage of maturity, when all the recommenders are already deployed in the recommender pool, the different recommenders analyze different aspects of the data, from demographic features to ratings, and provide an ensemble of recommendations to the users, based on different approaches and with varying degrees of personalization. The approach is novel in how it uses several techniques, from domain ontology to bin the items using NLP to achieve concept similarity, and then from there applies content-based, demographic-based, popularity-based and collaborative filtering approaches to attain the recommended items. The collaborative filtering employed are field-aware factorization machines which are the state-of-the-art in matrix factorization, which can easily include context-awareness. The aim is to provide a powerful and adaptable recommender system framework which can adapt to any domain, given the respective domain ontology, and can overcome cold-start issues by using an approach with 4 stages of maturity, which are subsequently entered when given thresholds are reached. In the following the structure of the paper is presented, with an explanation of every section. In the present section, the Introduction, an overview of the presented recommender system framework is provided as well as a literature review of the relevant works on the subject. In section 2, the framework and all its components are presented, from adopted technologies to used algorithms and techniques. A presentation of the architecture is given as well as a mock-up of the designed UI to provide the link between user and recommender system. In section 3, the technologies and techniques, mainly the ones central to the recommender system are better explained with some formulas being provided. In section 4, the recommender system is tested with a synthetic dataset with varying stages of maturity, to show how the recommender system evolves as the data changes. In section 5, conclusions are given as well as a brief discussion on future works.

## 1.1   Literature review

Recommender systems (RS) have been the focus of research for many years now, both on the algorithm side and on the applied side. The study of RS started in the beginning of the 1990s but it was in the last 15 years that research and the number of publications on the topic surged. Concerning our application, tourism, RS have been the focus of studies since, at least, the start of the 2000s with many publications having been made since then [1]–[15]. As for works that concern more an algorithmic approach without an explicit thematic application, several studies have been published on the different types of RS, from content-based approaches to collaborative filtering, as well as context-aware solutions, so called CARS [16]–[64].

Going into more detail regarding the tourism themed recommenders, it is relevant to give particular attention to ontology-based approaches. One of the more important examples concerning the present work is Moreno, A. et al. [8]. In this work, an ontology-based approach (SigTur/E-destination) is developed to get recommendations for tourism in the region of Tarragona. The developed approach begins with the definition of a tourism domain ontology, which describes the tourist activities in a hierarchy, and bins the activities according to a given taxonomy. The ontology is thus used to explicitly classify the activities to recommend among a predefined set of distinctive main concepts, which are used by the intelligent recommender system in its reasoning processes. The recommender than applies collaborative and content-based techniques to provide the recommendation. Another relevant work is that of García-Crespo, A. et al. [11], which proposes a semantic based expert system to provide recommendations in the tourist domain (Sem-Fit). The proposed system works based on the consumer's experience about recommendations provided by the system. Sem-Fit uses the experience point of view in order to apply fuzzy logic techniques to relating customer and hotel characteristics, represented by means of domain ontologies and affect grids. An early and interesting work that applies Bayesian networks to attain personalized recommendations for tourist attractions by Huang, Y. and Bian, L. [15] is also worth mentioning. This work is from 2009 and uses ontologies to classify different types of tourist attractions. It then uses a Bayesian network, to calculate the posterior probabilities of a given tourist's preferred activities and the traveler category he fits into. Other works on recommender system tourism applications could also be mentioned but instead one can mention three surveys done on this topic. First, one from 2014, Borràs, J. et al. [2] present a survey entitled "Intelligent tourism recommender systems". In this survey the various works in the state-of-the-art are analyzed and their different approaches concerning user interface, functionalities, recommendation techniques and use of AI techniques are presented. The second work that gives an overview on the topic is from Kzaz, L. et al. [3] from 2018. In this overview, the focus is essentially on recommender approaches and employed user and item data models. A third survey on this topic is given to us by Renjith, S. et al. [60] in a work titled "An extensive study on the evolution of context-aware personalized travel recommender systems". Herein, the authors start by defining the different recommender approaches that can be employed: content-based, collaborative, demographic-based, knowledge-based, hybrid, personalized and context-aware. The authors also go into detail on the different machine learning algorithms that are commonly employed, as well as the different employed metrics to evaluate the quality of the predictions. Finally, they present a table with many different works with the identification of whether or not they employ the previously mentioned techniques.

One of the aspects of the present work is that, as happens with some of the examples given above, it employs ontologies to organize and classify the items to be recommended in some way. Two works can also be mentioned concerning tourism domain ontologies, but in this case their formulation rather than their use. These works are by Ruíz-Martinez, J. et al. [65] and Barta, R. et al. [66] and they present different approaches to integrate and define tourism domain ontologies. In the latter work an approach is presented that shows how to cover the semantic space of tourism and be able to integrate different modularized ontologies. In the former, a strategy to automatically instantiate and populate a domain ontology by extracting semantic content from textual web documents. This work deals essentially with natural language processing and named entity recognition, which are some of the techniques also employed in this paper in terms of ontology population or, in other words, the classification of the different items to recommend according to the ontology.

Many other works should also be referenced, this time not necessarily linked to the tourism theme, but instead due to their focus on the algorithmic aspect or rather the recommendation strategy regardless of its field of application. One particular type of recommender system that is very much dominant in the literature in recent times is the context aware recommender system (CARS). The work by Kulkarni, S. et al. [32] gives us a review on the state-of-the-art techniques employed in context aware recommender systems. In this work the authors list the most common algorithmic approaches from bio-inspired algorithms to other common and less common machine learning algorithms and then enumerate the works that employed each type of solution. Another review study on context aware recommender systems is authored by Haruna, K. et al. [67]. In this work, the authors particularly emphasize the manner in which the contextual filtration is applied, for which there are three variants, pre-filtering, post-filtering and context modelling. The difference between each approach has to do with how context filtering is applied together with the process of recommendation. Hence, in pre-filtering the recommender filters the items prior to recommendation, while in post-filtering the opposite happens. In context modelling there is a more complex integration of the context filtering and the recommendations. The authors then go on to classify the different works in the literature according to this and other topics such as employed algorithms, etc. A third overview paper on the topic of CARS is the work by Raza, S. et al. [44]. In this work, the authors focus on the type of algorithms, the dimensionality reduction techniques, user modelling techniques and finally the evaluation metrics and datasets employed. Still focusing on CARS, a context-aware knowledge-based recommender system for movie showtimes called RecomMetz is presented in the work by Colombo-Mendoza, L. et al. [58]. In this work, the CARS developed has time awareness, crowd awareness and location awareness, as part of its context awareness composition. It is interesting to verify that its location awareness employs an exponential distance decay that discards items that are far away from the user. This sort of mechanism is also employed in the current work but with other goals. A last example on CARS is a genetic algorithm (GA) approach based on spatio-temporal aspects [68] by Linda, S. et al. Here, the interesting aspect is the inclusion of a GA to optimize the temporal weights of each individual while employing collaborative filtering for the recommendations.

Lately, one of the most studied techniques for recommender systems have been Factorization Machines (FM) [69]. In the present work, a field-aware version of this technique is employed, also known as an FFM. This technique is a kind of collaborative filtering method that gained some notoriety for solving click-through prediction rates [64], among other problems. Several versions exist of these FMs in the literature, with ensembles with deep neural networks [45], for example, being one of such versions. The value of FM is that they are more powerful than traditional matrix factorization techniques, being able to incorporate features and information such as implicit feedback. For these reasons, an FM, more specifically an FFM, is one of the recommenders employed in the proposed recommender system, constituting the collaborative filtering component of the proposed RS.

## 1.2 Description of the RS and field of application

The proposed RS in this work is to be applied in the tourism industry. More specifically, the project entails the creation of a recommender system to be used by hotel companies to recommend to their guests their vast lists of partners in the region. It is very common that large hotel companies have hundreds of partners offering products and most hotel guests are unaware of most of them. The partners usually offer a wide array of products, which need an ontology to be organized and better recommended. The proposed RS starts by having a Partner Management Platform (PMP) for the hotel's partners where they can manually introduce the items they want to be recommended in the RS. The PMP, which is essentially an interface of the Item DB, feeds the Domain Ontology which exists in a graph DB. The users are clients of the hotel that have checked-in, and they exist in the User DB, which houses not only demographic information but also user preferences which are collected and inferred by the RS. The RS interface is a web-app which is presented in a further section of the paper. In the following sections more detail is provided concerning the various components of the RS, starting with the presentation of the RS architecture in the following section.

## 2 Architecture and frameworks of the recommender system

The architecture of the RS can be essentially divided into 4 parts, the data repository, the context-aware subsystem, the recommender system per se and the user interface. In the following figure the architecture is presented with each of its subcomponents. An overview of each of the subcomponents is given in the following subsections.

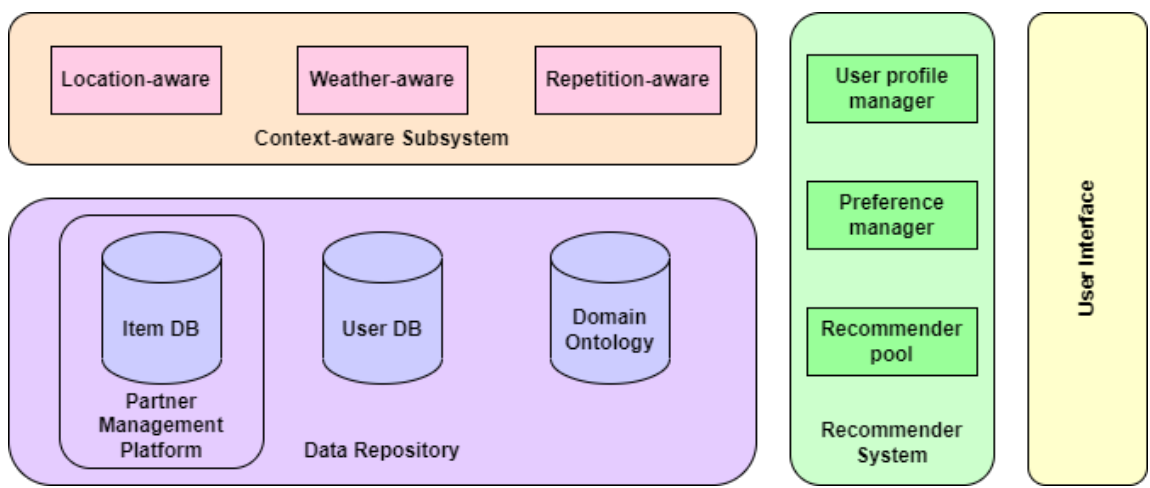

*Figure 1 Architecture of the RS.*

## 2.1 Data repository

The first element of the recommender system is its data repository, in the sense that this is where it starts, particularly with the Partner Management Platform (PMP). It is through this PMP that we have the introduction of the items, by the partners, to be recommended by the RS. In the PMP, the partners introduce the items alongside necessary description and keywords. This information introduced in the PMP is organized into an Item DB and later inserted into the domain ontology, which is later explained in detail.

Other than the PMP with its Item DB and the mentioned domain ontology, the data repository also has a User DB. This DB has both the demographic information collected from the users that check-in to the hotel, but also the preference vectors that are inferred and managed by the RS. The RS uses these two components of the user info to make predictions and to build different recommendation models based on demographic, content, and collaborative filtering techniques.

### 2.1.1   Domain ontology - Neo4j and automatic population of ontology

As for the domain ontology, the initial approach was to adopt the ontology presented in SigTur [8]. In addition, Neo4j ([www.neo4j.com](www.neo4j.com)), which is a graph DB, was chosen to house the ontology and to facilitate the automatic ontological extension with the items from the PMP. In the following figures, the original ontology is shown already inserted in a Neo4j graph.

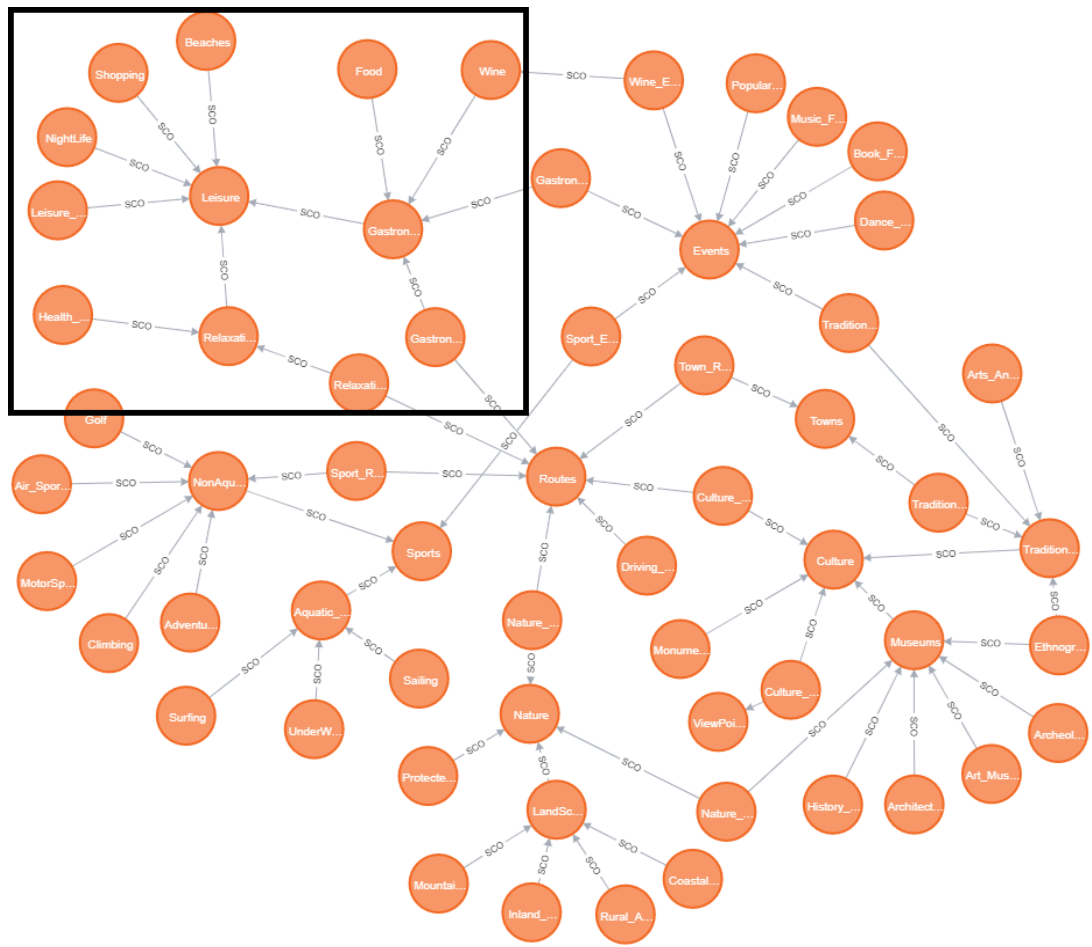

*Figure 2 Ontology inserted in Neo4j.*

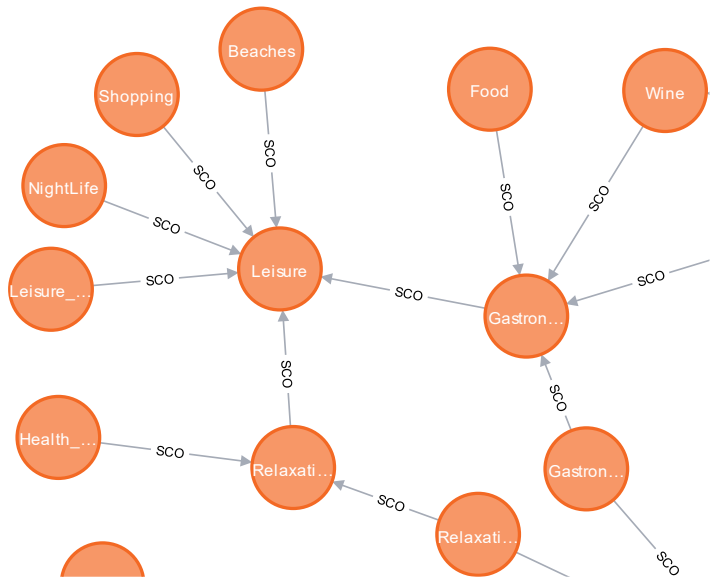

*Figure 3 Sample of the ontology (highlighted section in previous figure).*

The advantage of using the Neo4j framework is that it facilitates the automation of ontological extension. This ontological extension is achieved through the use of NLP techniques, such as named entity recognition and cosine similarity between semantic concepts, using the spaCy Python library integrated with Neo4j methods. These processes start with the insertion of the items from the PMP or the Item DB. These items are parsed and tokenized, using both the item descriptions and/or keywords. These parsed and tokenized items are then linked to the ontology by means of semantic similarity between its keywords and description with each of the ontological subclasses. The similarity scores above a given threshold originate a link between the item and that specific ontological subclass. This process that ends with concept similarity and starts with parsing, removal of stopwords and tokenization is performed with methods in the spaCy library. The concept similarity is performed using spaCy's vast pretrained word vectors. In addition, named entity recognition is also performed on the items, automatically linking a Wikipedia entry, if such entry exists. In Figure 4, a representation of the ontology after being extended with some items, via the described process. One can see the original nodes in orange, that belong to the ontology classes, some of which are now linked to grey nodes representing the items. The green nodes represent the Wikipedia page object when such an object was found. In Figure 5 a zoomed view of the highlighted zone in Figure 4 is shown. One can see two instances in which a Wikipedia page object was found from the Named Entity Recognition procedure. The items were linked to the ontology subclasses and one can observe that the links make sense in these cases, with driving an F1 racecar linked to "Motor Sports", and golf lessons and discounts on clubs linked to "Golf".

*Figure 4 Ontology extended with the addition of items.*

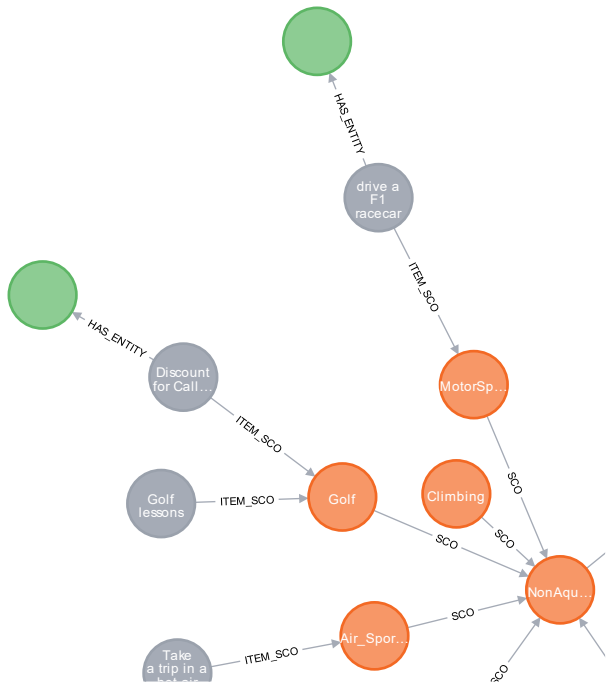

*Figure 5 Sample of the extended ontology (highlighted section in previous figure).*

The recommender system module then imports the extended ontology, both the classes and the items. It will use the extended ontology to give content-based recommendations.

## 2.2    Context-aware subsystem module

The context-aware subsystem module does item pre-filtering on the basis of three context submodules: location-aware, weather-aware and repetition-aware. In the case of the location-aware submodule, the objective is to filter out the hotel partners that are not located close by to a specific instance of the hotel. Since the hotel company can have a wide array of partners that may, in many cases, be close to one specific hotel but not to other hotels in other locations, such as local or regional partners that only provide services to the hotels in the area, a first contextual filtering phase is to apply location pre-filtering. Then we go on to the weather-aware submodule, where the ontological sub-classes are associated with a given fuzzy definition of when they make sense to be recommended, for example the beach ontology class or the outdoor sports ontology class would tend to be penalized with bad weather. Finally, a third module, which is very much novel, which is the repetition-aware module. Here, each ontological class would have a different elapsed time parameter that affects an inverse exponential penalization factor to mimic the repeatability of a given item. For example, one would probably be more adept to repeat a restaurant than a museum in the same week. So, different ontological classes have different factors that affect the inverse exponential function, that we may call the unwillingness to repeat function, which defines how soon a user may be willing to repeat a given item.

## 2.3 Recommender system module

The recommender system module is the main module as the name entails. This module is constituted by a user profile manager and a preference manager, besides the recommender pool. Concerning the recommender pool and the models that compose it, that is addressed in depth in Section 3 of this work. Here it suffices to say that the recommender pool is the set of different recommender models that provide user recommendations. The models create an ensemble, when more than one is active, that provides recommendations using different techniques and approaches.

As for the remainder of the recommender system module, the user profile and the preference manager, these two sub-modules manage the user related information, such as item ratings and other user feedback in the case of the former, while the latter manages the user preference vectors and propagates the user feedback on items to update the user preference vectors accordingly. The way this is done will become clearer in the next sections.

## 2.4 User interface – web app

The last component is the user interface, which in this case is a web app that connects to the recommender system module and other modules through a real-time and batch inference endpoints that connect to ML pipelines defined in Azure.

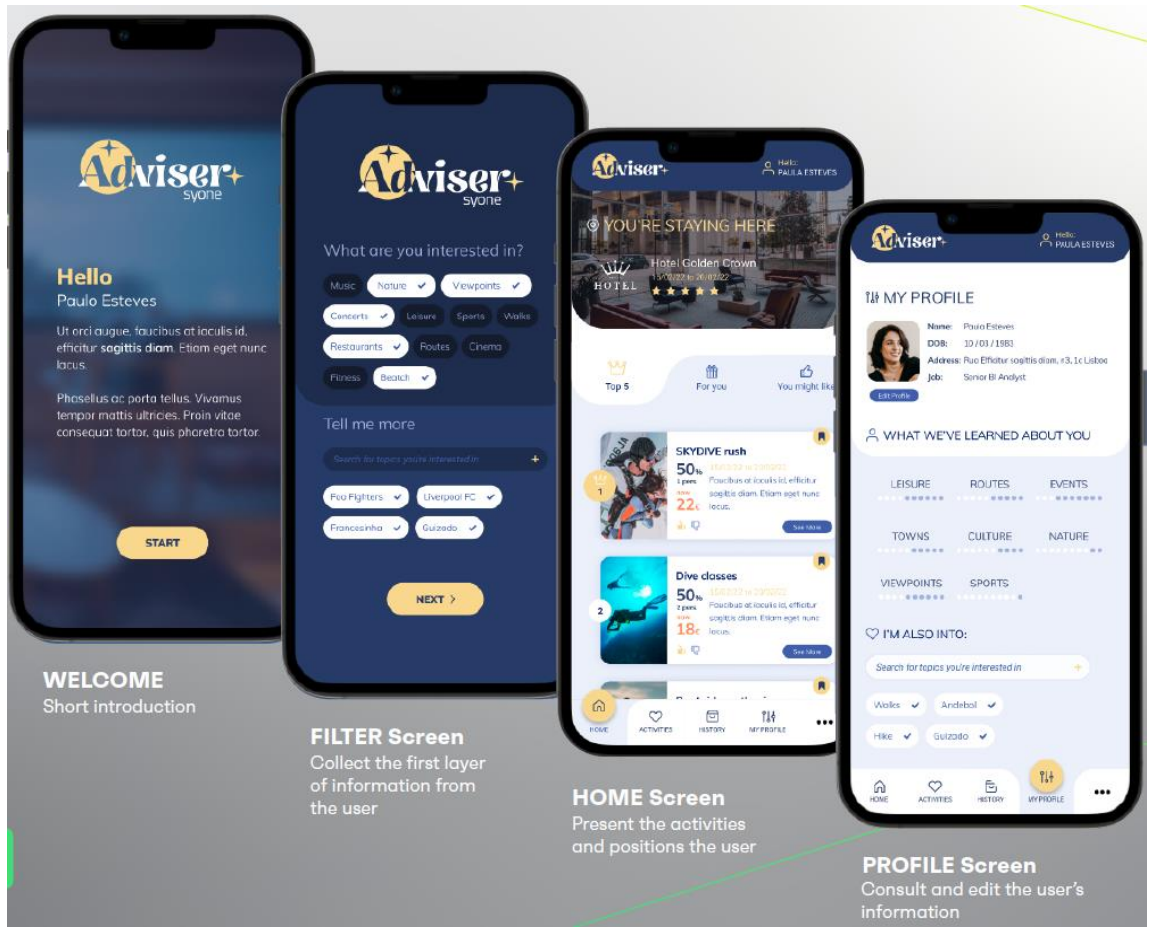

*Figure 6 App mockup showing the four main screens: welcome, preference definition, home and user profile.*

In the previous figure one can observe the four different screens the user sees during his App experience. The FILTER screen is only presented to the user on the first time he logs in and is, in essence, a series of check boxes where the user defines his preferences. These check boxes are used to give a first estimate on the user's preferences concerning the ontology classes. The user's choices define his preference vectors which then are used to make content-based recommendations. As for the HOME screen, it shows the different recommendations made to the user by the RS, here the user can bookmark items, book items or mark an item as "uninteresting". Finally, in the PROFILE screen, the user can observe his profile in terms preferences collected and inferred by the RS as well as demographic information, such as date of birth, nationality, etc.

The different interactions the user can have with the App and the consequent interactions between the App and the RS and back to the user are shown in Figure 7. In this figure one can see how these interactions cascade and what the user gets back from each action he undertakes. One can summarize the actions the user can take in the following:

- Logging in
- Preference input
- Viewing recommendations

- Item feedback
- Item booking
- Item rating

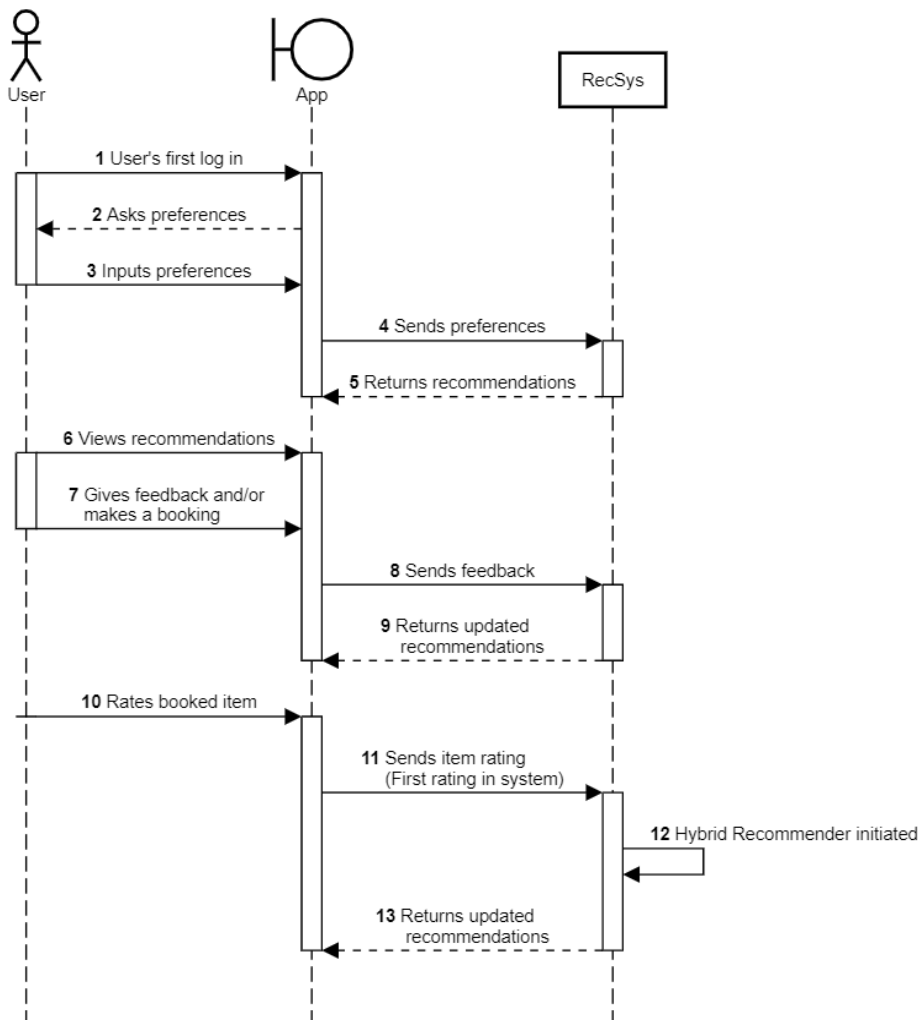

*Figure 7 User-App-RS interaction. User's various possible actions and respective interactions between the App and the RS.*

## 3    Recommenders and stages in RS

The recommender system module mentioned in the previous section is composed by three components: user profile manager, preference manager and recommender pool. The two former ones have already been covered, and in this Section, the latter will be explained in depth. The recommender pool is composed by four recommenders of different types: content-based, popularity-based, demographic-based and collaborative. These four recommenders are modeled with specific algorithms or employ specific techniques and they come into play in different phases of maturity of the RS. These phases of maturity concern amount of data, that is, number of users and rating density. Only after certain pre-specified values of users and rating density have been reached are some of these methods activated, or in other words, are some of the phases reached. In the following, the different phases and algorithms used are explained.

## 3.1 Phase 1

At the beginning, the RS is void of any ratings or users, and only items exist in the RS. When a new user logs in for the first time, in order for the RS to make any meaningful recommendation, some information has to be provided in the form of user preferences. This is, at this stage, the only way to overcome cold-start issues. The user's preferences, which are associated to the predetermined ontology are given and used to give content-based recommendations to the user. The user then will provide explicit and implicit feedback, in the form of booking items, bookmarking items or explicitly indicating they don't like the item. This feedback is then received by the RS who then uses the said feedback to update the user's preference vectors. This update originates new recommendations to the user.

### 3.1.1 Preference vectors

At the core of phase 1 are the user preference vectors. These preference vectors are ontology related and they are used to make content-based recommendations. There are three preference vectors per user:

- High-level preferences
- Low-level preferences
- Specific preferences

The high-level preferences are the ones the user identifies in the beginning and are associated with the ontological super-classes. These classes are the most abstract classes and lower in number. They are the first layer of ontological classes and are the ones that don't have a parent class and only child classes. Observing Figure 4, the *Sports* ontological class is an example of a high-level preference since there is no ontology class above it.

The low-level preferences are associated to the ontological classes that link directly to the items. These ontological classes are more specific, less abstract and in larger number. Observing Figure 4 and Figure 5, *Golf* is an example of a low-level preference, because two items link to it.

Finally, the specific preferences relate directly to the items, and is a vector that results from the other two higher-level preference vectors and the user's feedback on the items.

The way these vectors interact is explained in the following:

1. The user identifies the high-level preferences when he logs in for the first time. These preferences are propagated by way of vector multiplication with the low-level ontological preferences.
2. The low-level preferences are then propagated to the item level by way of vector multiplication as well, originating the specific preference vector. The items are ranked, and a subset of the highest ranked items are recommended to the user.
3. The user gives feedback on the recommendations by either bookmarking items, booking items or dismissing items. The feedback is propagated upwards to the higher-level preference vectors with different intensities. The low-level preference vector is strongly affected, while the high-level preference vector is less affected because it is higher upstream. This sort of "trickle-up" propagation of user feedback alters both high-level and low-level preference vectors with different magnitude.
4. New item recommendations are calculated, this time using both the high-level and low-level preference vectors to predict whether an item should be recommended or not. The prediction by each vector is weighed and aggregated originating an ensemble prediction using both high and low preference vectors. The items are ranked, and a subset of the highest ranked items are recommended to the user.
5. Repeat step 3.

### 3.1.2 Ontological content-based recommender

The content-based recommender is essentially vector multiplication between preference vectors and content vectors. Content vectors are binary vectors which map one preference level to the items content or to another preference vector content, while preference vectors show the intensity levels of preference for each ontological category.

In step 4, the high and low preference vectors multiply with their corresponding item content vector originating a content-based prediction. Both predictions are weighed and aggregated, and a subset of the highest ranked items is recommended to the user. After the user's feedback both preference vectors are updated according to the "trickle-up" propagation concept introduced above. Then, new recommendations are calculated with the new preference vectors.

### 3.2 Phase 2

If the user booked and used an item, he can then rate said item, which will kickstart the hybrid recommender composed by the initial content-based recommender and the new popularity-based appendix. This popularity-based recommender uses a so-called damped mean on every item so that little cardinality of ratings doesn't give an exaggerated edge of an item over another, such as an item with a single 5-star rating having a 5-star average.

$$Damped\ Mean_j = \frac{\sum_{i=1}^{n} r_{ji} + k \cdot \bar{\bar{r}}_G}{n + k}$$

Where $r_{ji}$ is item j's rating i, $k$ is the damping coefficient, $\bar{\bar{r}}_G$ is the global mean rating or some other default value, and $n$ is the number of reviews of item j.

### 3.2.1  Hybrid recommender (content-based + popularity-based)

The start of the hybrid recommender marks the start of phase 2. At this point in the RS, there aren't many users and there aren't many ratings. The lack in both mean that popularity-based, demographic-based or collaborative approaches are still of little use. As more users join and more ratings are given, other recommenders can become increasingly useful. As we reach a given threshold of user and rating numbers we can initiate the demographic-based recommender.

The way in which the hybrid recommender uses both recommenders is by cascading ensemble. That is, the popularity recommender pre-filters the items according to a rating threshold and then the content-based recommender recommends items that were not eliminated by the popularity recommender.

## 3.3  Phase 3

As more users are added to the RS, and as these users give feedback on recommended items, other types of recommenders can enter the recommender pool. A first set of threshold values for number of users and rating density is defined. When these thresholds are reached, phase 3 is initiated with yet another recommender being added: the demographic-based recommender.

### 3.3.1  Demographic-based recommender

The demographic-based recommender is composed by two ML algorithms. One clustering algorithm and one classification algorithm. The clustering algorithm has the purpose of identifying clusters of similar users based on their demographic features. The user's demographic features can be age, region/country, group composition, budget, academic degree, etc. These features can be a mix of numerical, ordinal and nominal features and so a clustering algorithm that can handle different data types is necessary. After the clustering has been performed, and the users are all organized in clusters, a classification algorithm is used to predict whether a user will enjoy each item based on the item feedback of other users in the same cluster.

For clustering, the algorithm employed was K-Prototypes, which works similarly to K-Means but can deal with mixed data types, particularly ordinal and nominal data. To define the clustering model, a knee region identifier is employed to automatically identify the optimal (or close to optimal) number of clusters. The clustering model is retrained from time to time when sufficient new users have been added since the last model fitting.

For classification a k-Nearest Neighbor algorithm, or kNN, was employed. Here, the users from the same cluster are used to predict whether a given user will enjoy the items, based on those users' feedback. The kNN uses a custom distance metric that takes into account both Jaccard and Manhattan distance metrics for the ordinal and nominal features. The kNN than weighs the opinion of the other users inversely proportional to their distance to the user to whom the predictions are being made. The predictions given by this algorithm are weighed and added to the predictions made by the hybrid recommender.

## 3.4 Phase 4

In phase 4, collaborative filtering is added to the pool. As it happens with phase 3, the entry into phase 4 takes place when thresholds of user cardinality and rating density are reached. Once this happens the collaborative filtering model is fitted and starts giving recommendations. The algorithm used for collaborative filtering is a Field-Aware Factorization Machine (FFM), which has already been introduced in Section 1. In the following sub-section, the FFM application is explained in more detail.

### 3.4.1 Collaborative filtering with Field-Aware Factorization Machines (FFM)

To use FFMs, a specific Python library (xLearn) is used and the data also has to be transformed into a specific format. A sample of a dataset in said format is shown in the following table.

*Table 1 Dataset in the FFM format where each column represents a feature, except for column 0 which represents the labels.*

|        | 0 | 1      | 2      | 3      | 4      | 5      | 6      | 7      | 8      | 9      |
|--------|---|--------|--------|--------|--------|--------|--------|--------|--------|--------|
| 0      | 0 | 0:1:1  | 1:2:1  | 2:3:1  | 3:4:1  | 4:5:1  | 5:6:1  | 6:7:1  | 7:8:1  | 8:9:1  |
| 1      | 1 | 0:10:1 | 1:2:1  | 2:11:1 | 3:4:1  | 4:5:1  | 5:6:1  | 6:12:1 | 7:13:1 | 8:14:1 |
| 2      | 0 | 0:15:1 | 1:16:1 | 2:3:1  | 3:4:1  | 4:17:1 | 5:6:1  | 6:18:1 | 7:19:1 | 8:20:1 |
| 3      | 1 | 0:15:1 | 1:2:1  | 2:21:1 | 3:22:1 | 4:17:1 | 5:6:1  | 6:23:1 | 7:8:1  | 8:24:1 |
| 4      | 1 | 0:10:1 | 1:16:1 | 2:3:1  | 3:4:1  | 4:17:1 | 5:25:1 | 6:23:1 | 7:26:1 | 8:27:1 |
| ...    | ... | ...  | ...    | ...    | ...    | ...    | ...    | ...    | ...    | ...    |
| 686422 | 1 | 0:1:1  | 1:2:1  | 2:3:1  | 3:4:1  | 4:17:1 | 5:25:1 | 6:23:1 | 7:8:1  | 8:37:1 |
| 686423 | 1 | 0:34:1 | 1:2:1  | 2:21:1 | 3:4:1  | 4:5:1  | 5:25:1 | 6:35:1 | 7:8:1  | 8:36:1 |
| 686424 | 1 | 0:10:1 | 1:16:1 | 2:3:1  | 3:4:1  | 4:17:1 | 5:25:1 | 6:18:1 | 7:8:1  | 8:24:1 |
| 686425 | 1 | 0:34:1 | 1:16:1 | 2:21:1 | 3:22:1 | 4:17:1 | 5:25:1 | 6:50:1 | 7:13:1 | 8:49:1 |
| 686426 | 1 | 0:15:1 | 1:2:1  | 2:3:1  | 3:4:1  | 4:17:1 | 5:6:1  | 6:23:1 | 7:8:1  | 8:44:1 |

This format is more complex than that for the Standard FM. This is due to the more complex information that is ingested by the FFM which uses information about the fields to define the latent vectors. That is, while in FMs each feature (field) has one latent vector, in FFMs this single representation is broken down into multiple latent vectors, one to represent each other field.

$$\hat{y}(x) := \omega_0 + \sum_{i=1}^{n} \omega_i x_i + \sum_{i=1}^{n} \sum_{j=i+1}^{n} \langle \mathbb{v}_i, \mathbb{v}_j \rangle x_i x_j$$

In the equation that represents the FM, which is shown above, the feature interactions represented by $\langle \mathbb{v}_i, \mathbb{v}_j \rangle$ would correspond to the following in our case scenario (user demographic features):

$$v_{male} \cdot v_{bluecollar} + v_{male} \cdot v_{lowbudget} + v_{male} \cdot v_{northeurope} + \cdots$$

That is, the male latent vector that multiplies with each other latent vector is the same. The idea behind FFM is that the weight of the male latent vector might not be the same when multiplying with the job latent vectors as they are with the budget latent vectors, and so on. Thus, in the FFM, the latent vectors are field-aware, which results in the following:

$$v_{male,job} \cdot v_{bluecollar,gender} + v_{male,budget} \cdot v_{lowbudget,gender} + v_{male,region} \cdot v_{northeurope,gender} + \cdots$$

Besides demographic features, as is shown in this example, the latent-vectors can also easily incorporate item features as well as contextual features and can thus integrate context-awareness in a deeper sense than simple contextual pre-filtering or post-filtering.

The FFM model represents the last phase addition to the recommender pool. The predictions attained from it are weighed and then aggregated with the predictions given by the other two, the hybrid and the demographic recommender. The weighs given to each recommender may be set to change over time so that it accompanies the maturity and complexity of each of the recommenders in the pool, thus giving progressively larger weight to the FFM as more users and more ratings are added to the system.

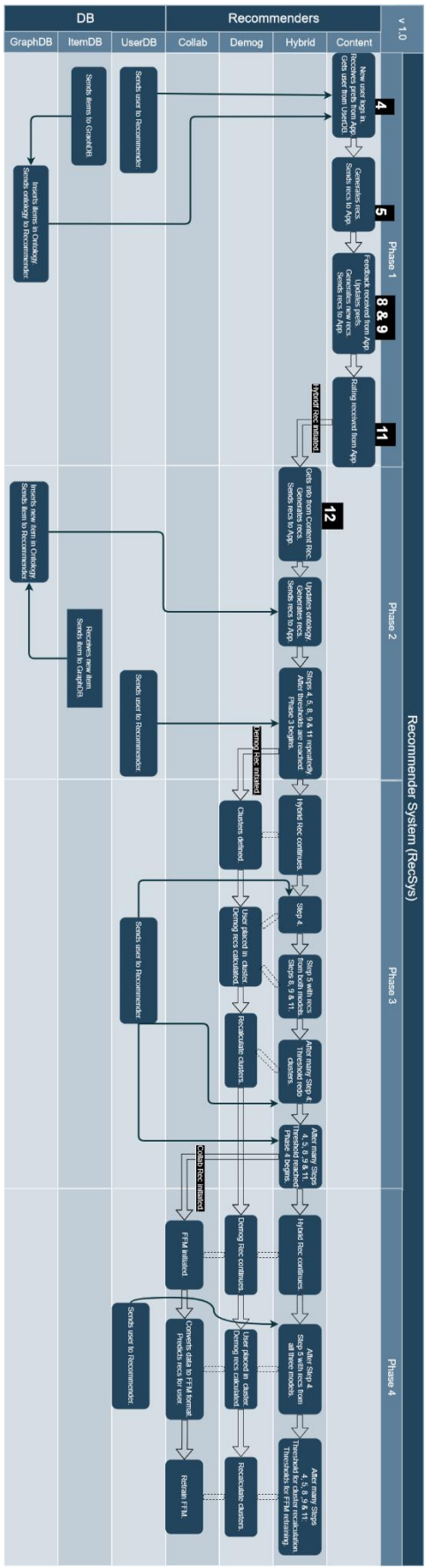

*Figure 8 Diagram of the various RS phases and interactions between RS and Data Repository (DB) components.*

# 4 Recommender system - Case study (CS) with synthetic data

One of the main challenges in designing the recommender system proposed in this work was the lack of data to perform any type of experiment or even just to aid and inspire in the definition of the algorithms to employ. The lack of data was absolute, both on the side of the items as on the side of the users and preferences. The main issue is the non-existence of a dataset with user demographic features and user preferences, since such a dataset would allow to overcome some of the cold-start issues as well as give some idea of the data schema to be adopted.

As a result, and since no public datasets were found that could overcome this hindrance, the decision was made to generate a synthetic dataset. The generated dataset was done so by using many different techniques from gaussian copulas to fuzzy logic. Further information on that work will be available in another paper by the author Camacho, VT. In the following sub-section, the synthetic data employed in this work's case study is presented.

Besides the synthetic data, a set of metrics was chosen to get an idea about the quality of the results from the recommenders. Traditional ML metrics are not always adequate for RS, mainly because, by principle, the objective of an RS is not to emulate exactly the choices of a given user since, if that were the case, there wouldn't be a need for an RS in the first place. In the metrics sub-section, the set of used metrics is presented.

The remainder of this section is applying the recommenders introduced in the previous section and testing them with different amounts of data which will attempt to emulate the data present at the different phases.

## 4.1 Synthetic data

In the work mentioned above, a methodology for the generation of synthetic datasets for recommender systems is presented, thus allowing to overcome the obstacle of not having quality data in sufficient amount (or even at all) readily available. The difficulties that are associated with this task are essentially the definition of a dataset with multiple datatypes, such as numerical (continuous), ordinal and nominal, and with different levels of correlation among the data, as well as the definition of user-ratings based on well-defined latent user preferences. To overcome this, a methodology was devised where several different techniques are employed in sequence to create the datasets concerning user characteristics, item properties, item categories and latent user preferences associated to user and item features, and as a result, a user-item sparse ratings matrix. The output of the methodology is:

1) Item dataset with item names and categories.

2) User dataset with user characteristics (demographic features).

3) User-item sparse ratings matrix.

4) Latent preferences and Multinomial Logit model to compare with the outputs of the Recommender System.

*4.1.1 Data Schema*

From the output presented above, we can see 4 DataFrames with different information. These DataFrames each have their own schema and have features from different data types. In the following, the created DataFrames are introduced:

- Demographic Features
- Preferences
- Item Features
- User Ratings

Going into more detail regarding the user demographic features DataFrame:

- Demographic Features:
    - User ID
    - Age
    - Gender
    - Job
    - Academic Degree
    - Budget
    - Country/Region
    - Group Composition
    - Accommodation

Concerning the type of feature, they can be divided essentially into three groups: numerical, categorical ordinal and categorical nominal. Concerning numerical and categorical ordinal features, we have the following:

- Numerical
    - Age – numerical (can be transformed into age bins)
- Ordinal:
    - Age bins = ['18-30','31-40', '41-50', '51-60', '60+']
    - Academic Degree = ['None', 'High School', 'Some College', 'College Degree']
    - Budget = ['Low', 'Mid', 'High']
    - Accommodation = ['Single', 'Double', 'Suite', 'Villa']

As for categorical nominal features, the following were modelled:

- Gender = ['Male', 'Female']
- Job = ['Blue Collar', 'White Collar']

- Country/Region = ['South Europe', 'North Europe', 'East Europe', 'North America', 'South America', 'Asia', 'Africa', 'Middle East']
- Group Composition = ['1 Adult', '2 Adults', '2 Adults + Child', 'Group of Friends']

### 4.1.2 Samples of the generated DataFrames

The resulting DataFrames (DF) can be used to train and test RS. In the case of the present work, they are used to simulate the different phases of data availability, thus testing the recommenders employed in each of the four phases. In the following, samples of the generated DFs are presented. The first sample shown is the User DF in Table 2. This DF is composed by the user demographic features and UserID. The demographic features are ordinal (Age, AcDeg, Budget, Accom) and nominal (Gender, Job, Region, GroupComp). The entire set of users created has cardinality of 100,000.

*Table 2 User DF composed by the demographic features of the users.*

| UserID | Age | AcDeg | Budget | Accom | Gender | Job | Region | GroupComp |
|---|---|---|---|---|---|---|---|---|
| 0 | 4 | 2 | 1 | 2 | Female | blue collar | North Europe | 2Adlt |
| 1 | 5 | 4 | 2 | 3 | Male | white collar | North Europe | GrpFriends |
| 2 | 3 | 3 | 2 | 2 | Female | blue collar | North Europe | 2Adlt+Child |
| 3 | 4 | 4 | 2 | 2 | Female | white collar | North Europe | 2Adlt+Child |
| 4 | 3 | 3 | 2 | 3 | Female | white collar | South Europe | 2Adlt |
| ... | ... | ... | ... | ... | ... | ... | ... | ... |
| 99995 | 4 | 4 | 2 | 2 | Female | white collar | North Europe | 2Adlt+Child |
| 99996 | 3 | 4 | 3 | 2 | Male | white collar | Asia | 2Adlt+Child |
| 99997 | 1 | 1 | 1 | 1 | Female | blue collar | South Europe | 2Adlt |
| 99998 | 1 | 3 | 1 | 2 | Female | blue collar | South Europe | 2Adlt+Child |
| 99999 | 4 | 3 | 2 | 2 | Male | blue collar | North America | 2Adlt+Child |

The second DF is the User-Preference DF which contains the latent preferences and is presented in Table 3. These latent preferences are related to the ontology classes. The latent preferences of each user were modeled through a multinomial logit model based on their demographic features. This DF shows the relative interest of a given user in a given preference category versus any other preference category. The values between different users are not comparable.

*Table 3 User-Preference DF containing the latent preferences from the Multinomial Logit model.*

| UserID | Beach | Relax | Shop | Nightlife | Theme park | Gastro | Sports | Culture | Nature | Events |
|---|---|---|---|---|---|---|---|---|---|---|
| *0* | 0.408 | 0.026 | 0.020 | 0.041 | 0.002 | 0.002 | 0.004 | 0.009 | 0.487 | 0.002 |
| *1* | 0.002 | 0.077 | 0.017 | 0.015 | 0.009 | 0.457 | 0.041 | 0.271 | 0.107 | 0.002 |
| *2* | 0.554 | 0.156 | 0.039 | 0.041 | 0.027 | 0.010 | 0.021 | 0.015 | 0.135 | 0.003 |
| *3* | 0.005 | 0.038 | 0.012 | 0.000 | 0.003 | 0.252 | 0.003 | 0.674 | 0.009 | 0.002 |
| *4* | 0.002 | 0.229 | 0.003 | 0.001 | 0.000 | 0.137 | 0.001 | 0.623 | 0.000 | 0.002 |
| *...* | ... | ... | ... | ... | ... | ... | ... | ... | ... | ... |
| *99995* | 0.003 | 0.106 | 0.202 | 0.000 | 0.020 | 0.115 | 0.005 | 0.202 | 0.337 | 0.010 |
| *99996* | 0.001 | 0.127 | 0.064 | 0.000 | 0.002 | 0.034 | 0.001 | 0.750 | 0.016 | 0.005 |
| *99997* | 0.050 | 0.285 | 0.030 | 0.337 | 0.110 | 0.006 | 0.091 | 0.019 | 0.015 | 0.057 |
| *99998* | 0.031 | 0.712 | 0.007 | 0.083 | 0.103 | 0.004 | 0.021 | 0.027 | 0.006 | 0.007 |
| *99999* | 0.005 | 0.880 | 0.064 | 0.000 | 0.035 | 0.000 | 0.009 | 0.003 | 0.002 | 0.003 |

The third DF sample presented is the Item DF in Table 4. Here a set of 29 items were included belonging to different categories which are the user latent preferences presented in the previous table.

*Table 4 Item DF with corresponding item category (ontology and latent preferences).*

| *itemID* | *Item Name* | *Category* |
|---|---|---|
| 0 | A service that offers you the opportunity to do bungee-jumping | ['Leisure', 'Sports', 'Routes', 'Events', 'Nature'] |
| 1 | A tavern that serves traditional food | ['Leisure', 'Events', 'Culture', 'Towns'] |
| 2 | Ancient history museum | ['Culture', 'ViewPoints', 'Events', 'Nature', 'Routes', 'Towns'] |
| 3 | Discount for Callaway clubs | ['Sports'] |
| 4 | Get a discount for Comic-Con | ['Sports'] |
| 5 | Get a free pint at the pub | ['Events', 'Leisure'] |
| 6 | Get a free pizza at Pizza Hut | ['Leisure'] |
| 7 | Get a voucher for Sephora | ['Leisure'] |
| 8 | Go shopping in our new mall | ['Leisure'] |
| 9 | Golf lessons | ['Sports', 'Leisure', 'Events'] |

| | |
|---|---|
| 10 | Great meals that are tasty | ['Leisure', 'Events'] |
| 11 | Medieval fair | ['Culture', 'Events', 'Nature', 'Towns'] |
| 12 | One day snorkeling with the fish | ['Sports', 'Leisure', 'Nature'] |
| 13 | One of the main nightclubs in the city | ['Culture', 'Events', 'Nature', 'Leisure', 'Routes', 'Towns'] |
| 14 | Rest and relaxation at the spa | ['Leisure', 'Routes'] |
| 15 | Surfing lessons | ['Sports'] |
| 16 | Take a trip in a hot-air balloon | ['Sports'] |
| 17 | Try go-karts with your friends | ['Sports'] |
| 18 | Try scubadiving | ['Sports'] |
| 19 | Try spearfishing with a pro | ['Sports'] |
| 20 | Watch a FC Porto match | ['Events', 'Sports'] |
| 21 | Watch a SL Benfica match | ['Events', 'Sports'] |
| 22 | Watch a Sporting CP match | ['Sports', 'Events'] |
| 23 | Watch a live concert of Mastodon | ['Events'] |
| 24 | Watch a live football match | ['Sports', 'Events'] |
| 25 | Watch a motogp race | ['Events', 'Sports'] |
| 26 | drive a F1 racecar | ['Sports'] |
| 27 | go to the spa | ['Leisure'] |
| 28 | visiting Disneyland | ['Leisure'] |

The last data sample is the result of an external product between the user preferences from the multinomial logit model and the item DF. The result is the input of a Fuzzy Inference System, which along with other implicit information on user and items returns the User-Item ratings DF, a sample of which is shown in Table 5.

*Table 5 User-Item ratings DF.*

| userId | 0 | 1 | 2 | 3 | 4 | 5 | … | 23 | 24 | 25 | 26 | 27 | 28 |
|---|---|---|---|---|---|---|---|---|---|---|---|---|---|
| 0 | 1.41 | 0.00 | 1.87 | 0.00 | 3.21 | 0.00 | … | 0.00 | 1.79 | 0.00 | 1.79 | 2.96 | 0.00 |
| 1 | 0.00 | 4.63 | 1.77 | 1.26 | 0.00 | 0.00 | … | 0.00 | 0.00 | 4.06 | 0.00 | 2.21 | 1.77 |
| 2 | 0.00 | 0.00 | 0.00 | 2.10 | 3.20 | 2.38 | … | 3.48 | 0.00 | 0.00 | 0.00 | 0.00 | 0.00 |
| 3 | 0.00 | 3.12 | 0.00 | 0.00 | 3.28 | 2.89 | … | 0.00 | 2.22 | 0.00 | 0.00 | 0.00 | 0.00 |
| 4 | 1.37 | 0.00 | 2.31 | 1.63 | 0.00 | 0.00 | … | 3.31 | 2.30 | 0.00 | 0.00 | 0.00 | 0.00 |
| … | … | … | … | … | … | … | | … | … | … | … | … | … |
| 99995 | 0.00 | 0.00 | 0.00 | 1.21 | 3.42 | 0.00 | … | 0.00 | 3.84 | 3.79 | 0.00 | 3.36 | 0.00 |
| 99996 | 1.46 | 0.00 | 0.00 | 0.00 | 2.31 | 0.00 | … | 2.31 | 0.00 | 0.00 | 0.00 | 0.00 | 1.39 |
| 99997 | 1.47 | 0.00 | 0.00 | 1.32 | 2.74 | 0.00 | …. | 0.00 | 0.00 | 2.29 | 0.00 | 0.00 | 0.00 |
| 99998 | 0.00 | 4.64 | 4.11 | 1.78 | 0.00 | 2.94 | … | 3.43 | 2.65 | 3.80 | 0.00 | 4.65 | 4.33 |
| 99999 | 0.00 | 3.54 | 3.06 | 0.00 | 4.07 | 2.65 | … | 0.00 | 3.07 | 3.51 | 2.46 | 3.50 | 2.61 |

## 4.2 Metrics

The metrics for a RS are not a trivial issue. Many works tend to use common ML metrics, such as classification metrics like precision, recall, accuracy, or regression metrics such as RMSE or MAE when the goal is to perform a regression on 1-5 ratings, for example. However, these metrics imply that the data available to us about user behavior is perfect, that is, users are aware of all the items they like and the ones they haven't tried aren't as relevant. If this were the case, no RS would be needed in the first place. The drawback of using this type of metrics is that it can encourage the recommender to make obvious recommendations in some cases, by penalizing wrong recommendations too much. In addition, these metrics do nothing to the tune of comparing recommenders based on how personalized its recommendations are, or how diversified.

Other metrics have been developed for RS in recent years that try to address these issues, some of which are presented in the following.

1. Mean Average Precision @ K and Mean Average Recall @ K

    As in more traditional machine learning, the dataset is split into training and test sets, and the test set is comprised of cases the learner did not train on and thus it is used to measure the model's ability to generalize with new data. In recommender systems, the same is done, and the output of a recommender system is usually a list of K recommendations for each user in the test set, and to produce those recommendations the recommender only trained on the items that user enjoyed in the training set. MAP@K (Mean Average Precision @ K) gives insight to how relevant the list of recommended items are, whereas MAR@K (Mean Average Recall @ K) gives insight to how well the recommender system is able to discover all the items the user has rated positively in the test set.

    In recommender systems, precision and recall are essentially the same as in machine learning:

    $$Precision = \frac{\# \ of \ relevant \ recommendations}{\# \ of \ items \ recommended}$$

    $$Recall = \frac{\# \ of \ relevant \ recommendations}{\# \ of \ relevant \ items}$$

    However, these metrics don't take ordering into account, and since the output of a recommender system is usually an ordered list, the metrics at cut-off k are introduced, MAP@K and MAR@K.

$$MAP@K = \frac{1}{|U|} \sum_{u=1}^{|U|} \frac{1}{\min(m,K)} \sum_{k=1}^{K} P_u(k) \cdot rel_u(k)$$

$$MAR@K = \frac{1}{|U|} \sum_{u=1}^{|U|} \frac{1}{m} \sum_{k=1}^{K} r_u(k) \cdot rel_u(k)$$

Where $U$ is the set of users in the test set, $m$ is the number of relevant items for user $u$, $P_u(k)$ and $r_u(k)$, are the precision@k and recall@k, respectively, and $rel_u(k)$ is a factor equal to 1 if the $k$ th item is relevant, and 0 otherwise.

2. Coverage

Coverage is the percentage of items on the training data that the recommender is able to recommend on a test set.

$$Coverage = \frac{I}{N} * 100\%$$

Where $I$ is the number of unique items the model recommends in the test data and $N$ is the total number of unique items in the training data.

3. Personalization

Personalization is the dissimilarity between users lists of recommendations. A high score indicates user lists are different between each other, while a low score indicates they are very similar. Similarity between recommendation lists is calculated via the cosine similarity between said lists and then by calculating the average of the upper triangle of the cosine similarity matrix (avgCosim). The personalization is then given by:

$$Personalization = 1 - avgCosim$$

4. Diversity

Diversity measures how different are the items being recommended to the user.

$$Diversity = 1 - ils$$

Where $ils$ corresponds to intra-list similarity, which is the average cosine similarity of all items in a list of recommendations. This calculation uses features of the recommended items (such as item metadata) to calculate the similarity. The feature matrix is indexed by the item id and includes one-hot-encoded features. If a recommender system is recommending lists of very similar items, the intra-list similarity will be high and conversely, the diversity will be low.

5. Novelty

Finally, novelty measures the capacity of recommender systems to propose novel and unexpected items which a user is unlikely to know about already. It uses the self-information of the recommended item, and it calculates the mean self-information per top-N recommended list and averages them over all users.

$$Novelty = \frac{1}{|U|} \sum_{u=1}^{|U|} \sum_{i=1}^{|N|} \frac{log_2\left(\frac{count(i)}{|U|}\right)}{|N|}$$

Where $U$ is the user list, $N$ is the top n-list and $count(i)$ is the number of users that have consumed the specific item.

## 4.3 CS with increasing data quantity

In this sub-section the previously presented datasets and the previously presented metrics are employed to test and evaluate the RS in its various phases. For this to work, the datasets will be gradually incremented, starting with very few users and no ratings, and ending with the full datasets. This process is meant to mimic the natural evolution of a RS, from initial cold-start conditions to thousands of users with thousands of reviews. In each phase different recommenders are employed as was already mentioned in previous sections.

### 4.3.1 CS in Phase 1

As mentioned previously, phase 1 is characterized by little number of users and no ratings. At this point, only content-based approaches are possible, and only if there is some input from the user concerning his preferences, which the RS asks when the user first logs in. Otherwise, the RS would be incapable of giving any recommendation short of a random context-filtered one. To mimic this first stage, 98 initial users are added to the RS. Each user inputs their HL preference vector related to Table 3, which the phase 1 content-based recommender uses to generate recommendations. Unlike in Table 3, the HL preference vector takes either 0 or 1 values and thus not conveying information on interest intensity. In the following tables, a sample of the 98 users and their respective HL vectors are shown.

*Table 6 High-level preferences of the users.*

| userId | ViewPoints | Nature | Towns | Culture | Events | Leisure | Routes | Sports |
|---|---|---|---|---|---|---|---|---|
| 1 | 0 | 0 | 0 | 0 | 0 | 1 | 0 | 0 |
| 2 | 1 | 0 | 1 | 0 | 0 | 1 | 1 | 0 |
| 3 | 0 | 0 | 0 | 0 | 0 | 1 | 0 | 0 |
| 4 | 0 | 0 | 0 | 0 | 0 | 1 | 1 | 1 |
| 5 | 0 | 0 | 1 | 0 | 0 | 0 | 0 | 0 |
| … | … | … | … | … | … | … | … | … |
| 94 | 0 | 0 | 0 | 0 | 0 | 1 | 0 | 0 |
| 95 | 0 | 0 | 0 | 0 | 0 | 1 | 0 | 0 |
| 96 | 1 | 0 | 1 | 0 | 0 | 1 | 1 | 0 |
| 97 | 0 | 0 | 1 | 0 | 0 | 0 | 0 | 0 |
| 98 | 1 | 0 | 1 | 1 | 0 | 0 | 1 | 0 |

The recommendations given by the RS for each user are in the following table. We can apply all previously presented metrics to these results, including MAP@K and MAR@K because we are aware of some ratings given by the users, present in the User-Item ratings DF which we can use for this purpose.

*Table 7 Sample of the recommendations given to the users by the content recommender.*

| userId | Recommendations |
|---|---|
| 1 | [(6, 'Get a free pizza at Pizza Hut'), (7, 'Get a voucher for Sephora'), (8, 'Go shopping in our new mall'), (27, 'go to the spa'), (28, 'visiting Disneyland')] |
| 2 | [(6, 'Get a free pizza at Pizza Hut'), (7, 'Get a voucher for Sephora'), (8, 'Go shopping in our new mall'), (14, 'Rest and relaxation at the spa'), (27, 'go to the spa')] |
| 3 | [(6, 'Get a free pizza at Pizza Hut'), (7, 'Get a voucher for Sephora'), (8, 'Go shopping in our new mall'), (27, 'go to the spa'), (28, 'visiting Disneyland')] |
| 4 | [(4, 'Get a discount for Comic-Con'), (6, 'Get a free pizza at Pizza Hut'), (7, 'Get a voucher for Sephora'), (8, 'Go shopping in our new mall'), (14, 'Rest and relaxation at the spa')] |
| 5 | [(11, 'Medieval fair'), (1, 'A tavern that serves traditional food'), (13, 'One of the main nightclubs in the city'), (2, 'Ancient history museum'), (0, 'A service that offers you the opportunity to do bungee-jumping')] |
| … | … |
| 94 | [(6, 'Get a free pizza at Pizza Hut'), (7, 'Get a voucher for Sephora'), (8, 'Go shopping in our new mall'), (27, 'go to the spa'), (28, 'visiting Disneyland')] |
| 95 | [(6, 'Get a free pizza at Pizza Hut'), (7, 'Get a voucher for Sephora'), (8, 'Go shopping in our new mall'), (27, 'go to the spa'), (28, 'visiting Disneyland')] |
| 96 | [(6, 'Get a free pizza at Pizza Hut'), (7, 'Get a voucher for Sephora'), (8, 'Go shopping in our new mall'), (14, 'Rest and relaxation at the spa'), (27, 'go to the spa')] |
| 97 | [(11, 'Medieval fair'), (1, 'A tavern that serves traditional food'), (13, 'One of the main nightclubs in the city'), (2, 'Ancient history museum'), (0, 'A service that offers you the opportunity to do bungee-jumping')] |
| 98 | [(2, 'Ancient history museum'), (11, 'Medieval fair'), (13, 'One of the main nightclubs in the city'), (1, 'A tavern that serves traditional food'), (14, 'Rest and relaxation at the spa')] |

*Table 8 Values for the various metrics on the content model recommendations.*

| MAP@K | MAR@K | Coverage | Personalization | Diversity HL | Diversity LL | Novelty |
|---|---|---|---|---|---|---|
| 0.092 | 0.092 | 0.55 | 0.51 | 0.21 | 0.76 | 0.66 |

We can see that mean average precision and mean average recall have the same value, the value at K is equal to 5, since the recommender recommends 5 items to each user. The two diversity values pertain to high level and low-level preferences showing how diverse are the recommendations in terms of recommending diverse items. It is expected for the high-level diversity to be lower than the low-level diversity since the content recommender makes recommendations based on high-level preferences of the users. Low-level preferences are linked ontologically to high-level preferences, but they are greater in variety, hence the same higl-level preference is linked to many low-level preferences, this justifies the larger value of Diversity LL compared to Diversity HL. Coverage, personalization and both diversities return values from 0 to 1, where 1 represents maximum coverage, personalization and diversity. The value for novelty can take any positive value, the greater the value the more unexpected recommendations are given based on popularity. In this study, the metric for novelty may not be very useful due to the relatively low cardinality of items and the fact that there are no less popular items per se, at least not very noticeably. In any case, these metrics are more useful in when used to compare different models.

### 4.3.2 CS in Phase 2

In phase 2 there are ratings in the system, although not enough users to feed the demographic-based recommender. In this phase we can simulate an RS state where there are 98 users and 64 ratings. The hybrid recommender is a hybridization of the initial content-based recommender with the new popularity-based recommender. The ratings are used to filter out items with average rating below a given threshold. Once again, the same metrics are applied, and the results are shown in the following table.

Table 9 Values for the various metrics on the hybrid model recommendations.

| MAP@K | MAR@K | Coverage | Personalization | Diversity HL | Diversity LL | Novelty |
|---|---|---|---|---|---|---|
| 0.219 | 0.219 | 0.17 | 1.11e-16 | 0.64 | 0.91 | 0.66 |

It is interesting to observe that the precision and recall have gone up, which makes sense because the items are now being filtered according to rating and higher rating items are more prone to having been liked by the users, at least the synthetic data was defined as such. The coverage has gone down, which makes sense since less items are being recommended due to filtering. Personalization has gone down since it now many users are being recommended the same items. Diversity has gone up; this can be due to recommending some items outside of the natural preference of the user due to ratings filtering. All in all, differences can be observed compared to the content-recommender, these differences make sense and seem to go towards an expected behavior by the recommender.

### 4.3.3 CS in Phase 3

In phase 3, enough users with ratings given have been introduced in the system to kickstart the demographic-based recommender. This recommender works by defining user clusters based on demographic features and then giving item recommendations based on the predictions of a kNN. This phase 3 recommender works together with the hybrid recommender from phase 2. In the following table, the metrics are applied, and the results shown. The number of users in this phase total 198, with 191 ratings.

Table 10 Values for the various metrics on the hybrid and demographic model recommendations.

| | MAP@K | MAR@K | Coverage | Personalization | Diversity HL | Diversity LL | Novelty |
|---|---|---|---|---|---|---|---|
| Hybrid | 0.178 | 0.178 | 0.34 | 0.07 | 0.64 | 0.91 | 0.66 |
| Demog | 0.151 | 0.151 | 0.72 | 0.57 | 0.63 | 0.90 | 0.66 |

We can see these results in a bar chart where a min max scaler has been applied. This basically shows which model wins in each category.

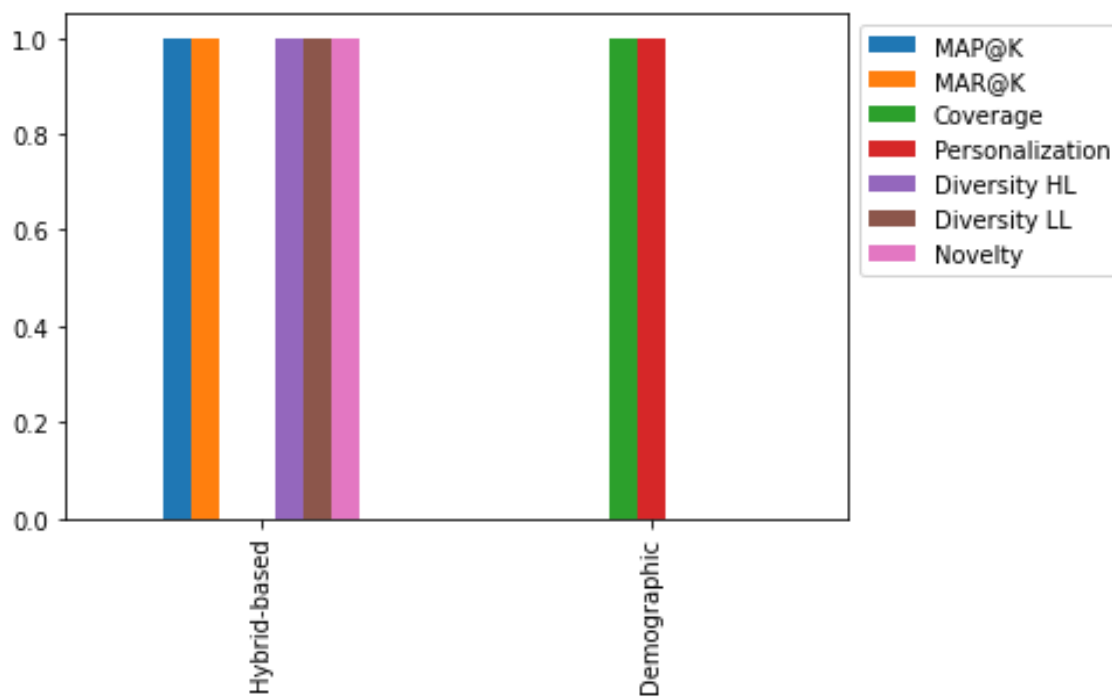

*Figure 9 Scaled metrics for both models.*

We can see that the hybrid model loses to the demographic model in coverage and personalization and has higher values in the other metrics. However, we can see that results are virtually equal in terms of Diversity and Novelty, and only on the Precision and Recall do we see larger values for the hybrid model, which are not that much higher. On the other hand, the demographic recommender has much larger personalization and coverage. Here we can see an increment by the demographic model compared to the hybrid model. This makes sense because the demographic model is more complex in how recommendations are given by finding similar users in terms of demographic features and then recommending similar items to the user on a more individual basis, whereas the hybrid model is again based on high level preferences.

*Table 11 Values for the various metrics on the hybrid phase 2 and hybrid phase 3 model recommendations.*

|  | MAP@K | MAR@K | Coverage | Personalization | Diversity HL | Diversity LL | Novelty |
|---|---|---|---|---|---|---|---|
| *Hybrid P2* | 0.219 | 0.219 | 0.17 | 1.11e-16 | 0.64 | 0.91 | 0.66 |
| *Hybrid P3* | 0.178 | 0.178 | 0.34 | 0.07 | 0.64 | 0.91 | 0.66 |

It is also interesting to compare the metrics between the hybrid in phase 2 and phase 3. We can see that most metrics remain similar with a slight decrease in precision and recall, which may be just random, a slight increase in personalization, and a rather large increase in coverage. This can be due to more items recommended and not filtered out due to poor ratings because of the existence of more users and ratings on items. It is interesting to see a variation of the metrics of the same recommender as the amount of data increases.

### 4.3.4 CS in Phase 4

Phase 4 starts when a given number of users and a given density of the user-item rating DF is achieved. When this happens, the final recommender is initiated. This recommender is the already mentioned FFM. In phase 4, the recommendations are, once again, the result of an ensemble of recommenders, the same one in phase 3 with the addition of the new FFM. The resulting metrics are once more applied to the recommendations and are shown in the following table. In this phase we have 250 users and 191 ratings.

*Table 12 Values for the various metrics on the hybrid, demographic and collaborative model recommendations.*

|        | MAP@K | MAR@K | Coverage | Personalization | Diversity HL | Diversity LL | Novelty |
|--------|-------|-------|----------|-----------------|--------------|--------------|---------|
| Hybrid | 0.158 | 0.158 | 0.34     | 0.06            | 0.64         | 0.91         | 0.66    |
| Demog  | 0.137 | 0.137 | 0.68     | 0.55            | 0.66         | 0.91         | 0.66    |
| Collab | 0.181 | 0.181 | 0.72     | 0.54            | 0.67         | 0.91         | 0.66    |

Comparing the recommenders, we can observe that the collaborative recommender, which was added in this later stage has high levels of personalization and coverage and achieves the highest values for precision and recall, compared to the other two models. The values for diversity are all similar at this stage, and novelty again doesn't provide useful information with this number of total items. In terms of precision and recall, coverage and personalization, the collaborative recommender gives us expected results which is relatively high values in these metrics. We can observe that each recommender brings different recommendations to the table with clear improvements in some metrics as the recommender system matures. It would be interesting to view this with a dataset comprising many more items and users. In the following figure we can see the metrics in a scaled graph.

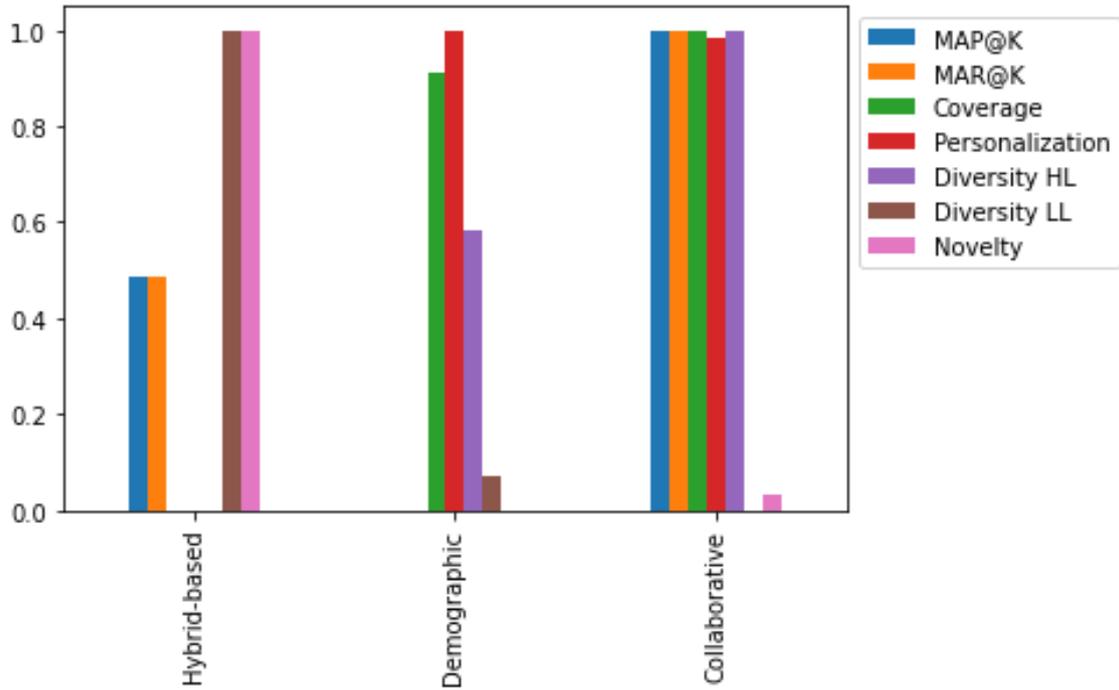

*Figure 10 Scaled metrics for all three models*

As said, we observe that the collaborative metrics are good in comparison to the other two, however, the collaborative model is only useful when the recommender system has seen sufficient data. The metrics for the other two are not as high but they don't suffer so much from cold-start issues. We can see that between the demographic and the hybrid models there is a trade-off in metrics. We had already seen this in the previous phase.

*Table 13 Values for the various metrics on the phase 1, phase 2 and phase 3 model recommendations of hybrid and demographic models.*

|          | MAP@K | MAR@K | Coverage | Personalization | Diversity HL | Diversity LL | Novelty |
|----------|-------|-------|----------|-----------------|--------------|--------------|---------|
| Hybrid P2 | 0.219 | 0.219 | 0.17 | 1.11e-16 | 0.64 | 0.91 | 0.66 |
| Hybrid P3 | 0.178 | 0.178 | 0.34 | 0.07 | 0.64 | 0.91 | 0.66 |
| Hybrid P4 | 0.158 | 0.158 | 0.34 | 0.06 | 0.64 | 0.91 | 0.66 |
| Demog P3 | 0.151 | 0.151 | 0.72 | 0.57 | 0.63 | 0.90 | 0.66 |
| Demog P4 | 0.137 | 0.137 | 0.68 | 0.55 | 0.66 | 0.91 | 0.66 |

Here we can see a comparison between the metrics of the different models along each phase, we can see a slight decrease of precision and recall in the evolving phases for hybrid and demographic models, but this might have to do with insufficient ratings being added between phase 3 and phase 4, which are important for the demographic recommender. With a further increase in data, we can see further differences in the metrics. Feeding the recommender system with 1000 users and 883 ratings, we attain the following results.

*Table 14 Values for the various metrics on the hybrid, demographic and collaborative model recommendations, in the case of 250 users and 191 ratings as well as 1000 users and 883 ratings.*

|  | MAP@K | MAR@K | Coverage | Personalization | Diversity HL | Diversity LL | Novelty |
|---|---|---|---|---|---|---|---|
| *Hybrid* | 0.158 | 0.158 | 0.34 | 0.06 | 0.64 | 0.91 | 0.66 |
| *Demog* | 0.137 | 0.137 | 0.68 | 0.55 | 0.66 | 0.91 | 0.66 |
| *Collab* | 0.181 | 0.181 | 0.72 | 0.54 | 0.67 | 0.91 | 0.66 |
| *Hybrid 1000* | 0.088 | 0.088 | 0.28 | 0.19 | 0.69 | 0.89 | 0.66 |
| *Demog 1000* | 0.128 | 0.128 | 0.97 | 0.59 | 0.48 | 0.89 | 0.66 |
| *Collab 1000* | 0.119 | 0.119 | 0.79 | 0.61 | 0.52 | 0.89 | 0.66 |

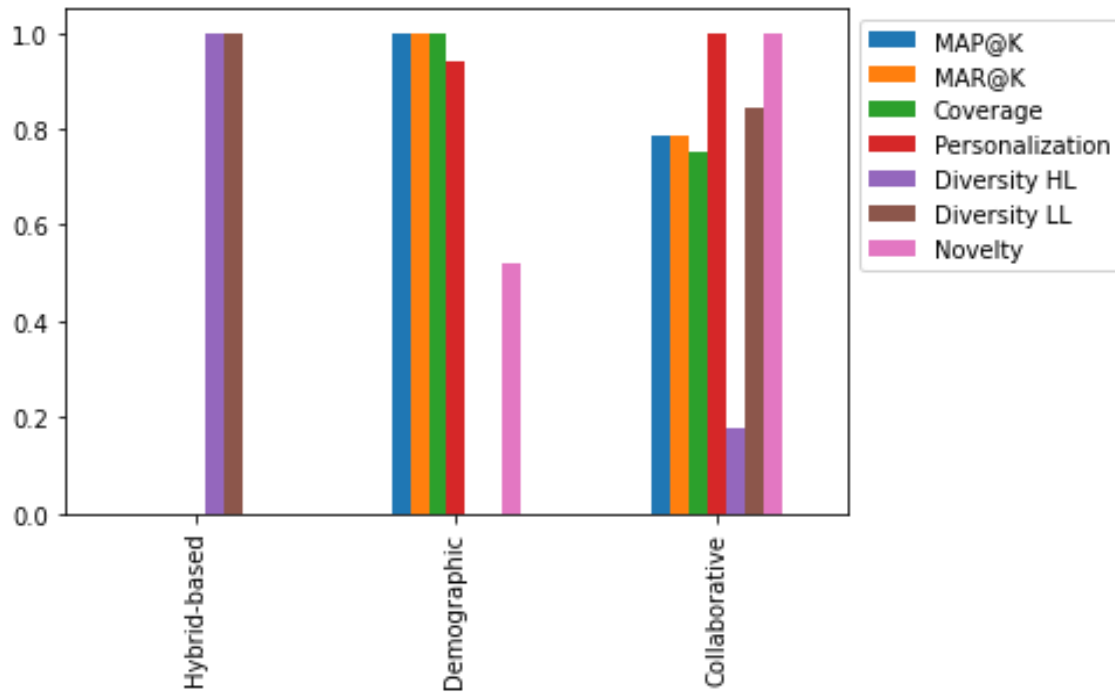

*Figure 11 Scaled metrics for all three models.*

We can see that the metrics are qualitatively similar to the case before with less users and ratings. Still the number of ratings is low, there is not a lot of rating density, which particularly penalizes the collaborative model. Nonetheless, we can observe that the collaborative model is the one that offers more personalization, which increased for all models with the increment in users and ratings. Coverage also increased heavily for the demographic model while only increasing slightly for the collaborative model. As for precision and recall, the demographic model maintains the metric with only a slight decrease while the hybrid and collaborative model saw a rather significant decrease. In regard to the collaborative model this might have to do with the low density in ratings. All in all we see that the demographic and collaborative models clearly become more dominant and useful as more data is added to the RS. The phases also make sense, by having the collaborative model initiate after all others have been initiated, since the collaborative model is very sensitive to rating density, while the demographic model is more robust in that sense. The hybrid model by this phase has clearly been passed by the two other models in most metrics which is exactly what would be expected.

## 5 Conclusion and future works

In this work an ontology-based context aware recommender system application for tourism was presented where different recommenders are used at different stages of maturity of the recommender system. The novel aspect is the evolution of the recommender system with different types of recommenders entering the recommendation pool as the system's maturity evolves. The ontology extension of the recommender system allows items to be binned and recommended to users based on user preference vectors with different degrees of detail that link to the item ontology. These preference vectors will be ever changing based on user feedback, while other recommenders based on demographic features and field-aware factorization machines join the pool as data increases.

Along this work, the RS was presented and ultimately tested with synthetic data mimicking different stages of maturity. One could observe that at each new phase the new recommenders added value as observed from the comparison between the different adopted metrics, which were MAP@K, MAR@K, Coverage, Personalization, Diversity HL, Diversity LL and finally Novelty. These metrics are the state of the art for Recommender Systems because they attempt to go beyond the usual metrics adopted in ML, which don't always have much meaning in RS. The results obtained were expected where Collaborative and Demographic approaches essentially brought more personalization and coverage to the table. However, the full extent of differences between recommenders could not be captured mainly due to the relatively low cardinality of items being offered, only 29.

Future works would entail a broader analysis with more items, and also context-aware data which was not tested at this instance. Nonetheless, the context-aware would be essentially pre-filtering which would not be of much interest regarding the results concerning the metrics.

## 6 Acknowledgements

The present paper was developed in the context of the PMP project – Partnership Management Platform, code LISBOA-01-0247-FEDER-045411, co-financed by LISBOA 2020 and Portugal 2020 through the European Regional Development Fund.